
\mag=\magstep1
\documentstyle{amsppt}
\topmatter
\title
Schur-Weyl Reciprocity for the Hecke Algebra \\
of $(\Bbb Z/r\Bbb Z)\wr \frak S_n$
\endtitle
\author
SUSUMU ARIKI, TOMOHIDE TERASOMA, \\
AND \\
HIROFUMI YAMADA
\endauthor

\affil
Susumu Ariki,\\
Division of Mathematics,\\
Tokyo University of Mercantile Marine,\\
Etchujima, Koto-ku, Tokyo 135, Japan \\
                   \\
Tomohide Terasoma, \\
Department of Mathematics, \\
Tokyo Metropolitan University, \\
Minami-Ohsawa, Hachioji, Tokyo 192-03, Japan \\
                 \\
Hirofumi Yamada, \\
Department of Mathematics, \\
Tokyo Metropolitan University, \\
Minami-Ohsawa, Hachioji, Tokyo 192-03, Japan
\endaffil






\endtopmatter
\document

\heading
Introduction
\endheading

The purpose of this paper is to give a reciprocity between
$U_q(h)$ and $\Cal H_{n,r}$, the Hecke algebra of
$(\Bbb Z / r\Bbb Z)\wr \frak S_n$ introduced by  Ariki and Koike [1].

The Schur-Weyl reciprocity was originally discovered for $GL(m)$ and
$\frak S_n$ [17, p.130]. This is the first example of dual pairs and
has been generalized to various pairs of groups and algebras.
Jimbo [10] proved a $q$-analogue of the original reciprocity, namely that
between $U_q(gl_m)$ and the Hecke algebra $\Cal H_n$ of type $A$.
A.Ram [14] utilizes the reciprocity to obtain a character formula of
$\Cal H_n$.

Let $K=\Bbb Q(q,u_1,\dots ,u_r)$ be the field of rational funcitons in
variables $q,$\linebreak
$ u_1,\dots ,u_r$. We adopt $K$ as the base field for both
the quantized universal enveloping algebra $U_q(gl_r)$ and the Hecke
algebra $\Cal H_n$.

We denote by $U_q(h)$ the $K$-subalgebra of $U_q(gl_r)$
generated by $q^{E_{ii}}\;$'s $(1\le i \le r)$. In this paper, we show that
the commutant of $U_q(h)$ in $End((K^r)^{\otimes n})$ is isomorphic
to a quotient of $\Cal H_{n,r}$. We also determine the irreducible
decomposition of $(K^r)^{\otimes n}$ under the action of $\Cal H_{n,r}$.
As a consequence, we obtain the reciprocity for $U_q(h)$ and
$\Cal H_{n,r}$.

Let us review the classical Schur-Weyl reciprocity in a modified sense, i.e.,
that between $U(\frak g)$ and $\frak S_{n,r}$. Here, $U(\frak g)$ denotes the
universal enveloping algebra of $\frak g=gl_{m_1}\oplus\cdots\oplus gl_{m_r}$,
and $\frak S_{n,r}$ is the group consisting of $n\times n$ permutation
matrices whose nonzero entries are $r$-th roots of unity. The group
$\frak S_{n,r}$ is generated by $\frak S_n$ and
$$
s_1=e^{2\pi\sqrt{-1}/r}E_{11}+E_{22}+\cdots+E_{nn}.
$$
The vector representation of $\frak g$ is $\Bbb C^m=
\Bbb C^{m_1}\oplus\cdots\oplus
\Bbb C^{m_r}$, which has the standard basis $v^i_j$ $(1\le i\le r,
1\le j\le m_i)$.
On $(\Bbb C^m)^{\otimes n}$, $\frak S_n$ acts by permuting components of
the tensor product. The action of $gl_m$ on $(\Bbb C^m)^{\otimes n}$
is infinitesimally diagonal. We can extend the action of $\frak S_n$
to that of $\frak S_{n,r}$ by letting $s_1$ act on $(\Bbb C^m)^{\otimes n}$
by
$$
s_1(v^{i_1}_{j_1}\otimes\cdots\otimes v^{i_n}_{j_n})=
e^{2\pi\sqrt{-1}i_1/r}v^{i_1}_{j_1}\otimes\cdots\otimes v^{i_n}_{j_n}\;.
$$

Since $U(\frak g)$ is a subalgebra of $U(gl_m)$, we naturally have the
action of $U(\frak g)$ on $(\Bbb C^m)^{\otimes n}$ through that of
$U(gl_m)$ on it.
Thereby, $(\Bbb C^m)^{\otimes n}$ is a $U(\frak g)\times
\frak S_{n,r}$-module.
It is well known that both of the irreducible representations of
$\frak S_{n,r}$ and $U(\frak g)$ occurring in $(\Bbb C^m)^{\otimes n}$ are
indexed by the set $\varLambda_{m_1,\cdots,m_r}(n)$ of r-tuples
$\underline{\lambda}=(\lambda^{(1)},\cdots,\lambda^{(r)})$ of
Young diagrams with $\sum_{i=1}^r\mid\lambda^{(i)}\mid=n$, and
$l(\lambda^{(i)})\le m_i$ for $i=1,\cdots,r$ [11].
The irreducible representation space corresponding to
$\underline{\lambda}\in\varLambda_{m_1,\cdots,m_r}(n)$ is denoted by
$W_{\underline{\lambda}}$ for $U(\frak g)$ and $V_{\underline{\lambda}}$ for
$\frak S_{n,r}$, respectively. Then we actually have,
$$
(\Bbb C^m)^{\otimes n}=\bigoplus_{\underline{\lambda}\in
\varLambda_{m_1,\cdots,m_r}(n)}W_{\underline{\lambda}}\otimes
V_{\underline{\lambda}}
$$
as a $U(\frak g)\times \frak S_{n,r}$-module. As a consequence, the each
image of $U(\frak g)$ and of the group ring $\Bbb C\frak S_{n,r}$ in
$End_{\Bbb C}((\Bbb C^m)^{\otimes n})$ is the full centralizer algebra of
the other.
The same situation also appears in a natural setting for finite fields,
which will be explained in appendix.

We will give a $q$-analogue of the above story for the special case that
$m_i=1$ for $i=1,\cdots,r$. It is an interesting problem to establish
a $q$-analogue for the general case.

\heading
\S 1 Preliminaries
\endheading

Let $U_q(\frak g)=U_q(gl_r)$ be the
quantized universal enveloping algebra of $\frak g = gl_r$ over $K=$
$\Bbb Q(q,u_1,\dots,u_r)$, defined by the following generators and
relations ([10]):
$$
\text{ Generators: }
\cases
&q^{\pm \epsilon_i}\; (1 \leq i \leq r), \\
&e_i\;(1 \leq i < r),\\
&f_i\;(1 \leq i < r).
\endcases
$$
$$
\gather
\text{ Relations: }
q^{\epsilon_i}q^{-\epsilon_i} =
q^{-\epsilon_i}q^{\epsilon_i} = 1 \;,
q^{\epsilon_i}q^{\epsilon_j} =
q^{\epsilon_j}q^{\epsilon_i} \;, \\
q^{\epsilon_i}e_jq^{-\epsilon_i} =\cases q^{-1}e_j \;&(j = i-1) \\
qe_j \;&(j = i) \\
e_j\; &(\text{ otherwise }),
\endcases  \\
q^{\epsilon_i}f_jq^{-\epsilon_i} =\cases qf_j \;&(j = i-1) \\
q^{-1}f_j \;&(j = i) \\
f_j\; &(\text{ otherwise }),
\endcases \\
e_if_j -f_je_i  = \delta_{ij}\frac{q^{\epsilon_i-\epsilon_{i+1}}
-q^{-\epsilon_i+\epsilon_{i+1}}}{q-q^{-1}} \;,
\\
e_{i\pm 1}e_i^2-(q+q^{-1})e_ie_{i\pm 1}e_i +e_i^2e_{i\pm 1} = 0 \;, \\
f_{i\pm 1}f_i^2-(q+q^{-1})f_if_{i\pm 1}f_i +f_i^2f_{i\pm 1} = 0 \;, \\
e_ie_j=e_je_i,\quad f_if_j=f_jf_i \quad(i>j+1).
\endgather
$$
It is well-known that $U_q(\frak g)$
has a Hopf algebra structure with coproduct $\Delta :
U_q(\frak g)\to  U_q(\frak g)\otimes U_q(\frak g)$
defined by
$$
\align
\Delta (q^{\pm\epsilon_i}) &=
q^{\pm\epsilon_i}\otimes q^{\pm\epsilon_i} \;,
\\
\Delta (e_i) &= e_i \otimes 1 + q^{\epsilon_i-\epsilon_{i+1}}\otimes e_i \;,
\\
\Delta (f_i) &=  f_i\otimes q^{-\epsilon_i+\epsilon_{i+1}} +
1 \otimes f_i \;.
\\
\endalign
$$
On the vector space $V = K^r$, with the standard basis $v_i$ ($1\leq i
\leq r$), an action $\rho$ of $U_q(\frak g)$ is defined by
$$
\align
\rho (e_i)v_j &= \cases
v_{j-1} \; &(j = i+1) \\
0 &(j \neq i+1), \\
\endcases
\\
\rho (f_i)v_j & = \cases
v_{j+1} \; &(j = i) \\
0 &(j \neq i), \\
\endcases
\\
\rho (q^{\epsilon_i})v_j & = \cases
qv_{j} \; &(j = i) \\
v_j &(j \neq i). \\
\endcases
\\
\endalign
$$
This is called the natural representation of $U_q(\frak g)$.
Put $\Delta^{(2)} =\Delta$ and
$\Delta^{(k)} = (\Delta^{(k-1)}\otimes id)\circ \Delta$
for $k \geq 3$. By using $\Delta^{(n)}$ such obtained,
one can make $V^{\otimes n}$ into a $U_q(\frak g)$-module:
$$
\rho(x)(v_{i_1}\otimes\cdots\otimes v_{i_n})=
\Delta^{(n)}(x)(v_{i_1}\otimes\cdots\otimes v_{i_n})
\quad(\text{for\;}x\in U_q(\frak g)).
$$
The action is also denoted by $\rho$.
We let $U_q(h)$ be the $K$-subalgebra of $U_q(\frak g)$ generated by
$q^{\pm \epsilon_i}$ ($1 \leq i \leq r$).

Denote by $\Cal H_n$ the Hecke algebra of the
symmetric group $\frak S_n$.
More precisely, $\Cal H_n$ is the $K$-algebra defined by the following
generators and relations:
$$
\align
\text{ Generators: }&
g_2, \cdots ,g_n \;.
\\
\text{ Relations: }&
(g_i-q)(g_i + q^{-1}) = 0
\; (2 \leq i \leq n),\\
& g_ig_{i+1}g_i =
g_{i+1}g_{i}g_{i+1}
\; (2 \leq i < n),
\\
 & g_ig_j = g_jg_i
\; (i>j+1).
\endalign
$$
The algebra $\Cal H_n$ also acts on $V^{\otimes n}$ by
$$
\sigma (g_k) = id^{\otimes (k-2)}\otimes \check R
\otimes id^{\otimes (n-k)},
$$
where $\check R \in End_K(V \otimes V)$ is defined by
$$
\check R (v_{i}\otimes  v_{j}) =\cases
qv_{i}\otimes  v_{j}\; &(i = j)  \\
v_{j}\otimes v_{i}
\; &(i> j)  \\
v_{j}\otimes v_{i}
+(q-q^{-1})
v_{i}\otimes v_{j}
\; &(i < j).  \\
\endcases
$$
We will denote $T_k = \rho (g_k)$ for $2\le k\le n$.
A $q$-analogue of the Schur-Weyl reciprocity due to Jimbo asserts
that each of $\rho (U_q(\frak g))$ and $\sigma (\Cal H_n)$ is
the full centralizer algebra of the other in
$End_K(V^{\otimes n})$.

We now recall the definition and properties of the Hecke algebra
$\Cal H_{n,r}$ for a positive integer $r$.
For further discussions about $\Cal H_{n,r}$, readers may refer to [1].
The Hecke algebra $\Cal H_{n,r}$ is the $K$-algebra defined by
generators and relations as follows:
$$
\align
\text{ Generators: }&
g_1,g_2, \cdots ,g_n \;.
\\
\text{ Relations: }&
(g_1 -u_1)\cdots (g_1 - u_r) = 0 \;,
\\
&
(g_i-q)(g_i + q^{-1}) = 0
\; (2 \leq i \leq n),\\
& g_1g_{2}g_1g_2 =
g_{2}g_1g_{2}g_{1} \;,
\\
& g_ig_{i+1}g_i =
g_{i+1}g_{i}g_{i+1}
\; (2 \leq i < n),
\\
 & g_ig_j = g_jg_i
\; (i>j+1).
\endalign
$$
Note that our $\Cal H_{n,r}$ is isomorphic to Ariki-Koike's when
$q$ is replaced by $q^2$ in [1].

We define $t_j$ $(j=1, \cdots ,n)$ recursively by $t_1 = g_1$,
$t_j = g_j t_{j-1}g_j$ ($j \geq 2$) and $\Cal T_{n,r}$ to
be the $K$-subalgbra of $\Cal H_{n,r}$ generated by these elements.

For an $r$-tuple of Young diagrams $\underline \lambda$
$= (\lambda^{(1)},\cdots ,\lambda^{(r)})$ with size \linebreak
$\sum_{i=1}^r \mid \lambda^{(i)} \mid = n$, a tableau
$\underline S$ $= (S^{(1)}, \cdots , S^{(r)})$
of shape $\underline \lambda$ is said to be standard if
each $j$ ($1\leq j \leq n$) occurs exactly once and each $S^{(i)}$
($1 \leq i \leq r$) is such a tableau that entries in each column
are increasing from top to bottom and in each row from left to right.
If $i$ is located in the intersection of the $l$-th row and the $m$-th
column of $S^{(p)}$, then we write $\tau(\underline S;i)=p$ and
$c(\underline S;i)= m-l$.
For each standard tableau $\underline S$ we associate a character
of $\Cal T_{n,r}$ by
$$
\varphi_{\underline S}(t_i) = u_{\tau(\underline S;i)}q^{2c(\underline S;i)}
\quad(1\le i\le r).
$$
\proclaim{Proposition 1.1 ([1])}
\roster
\item
The algebra $\Cal T_{n,r}$ is commutative and semi-simple.
\item
The complete set of irreducible representations of $\Cal T_{n,r}$
is
$$
\{ \varphi_{\underline S} \mid \underline S \text{\;is standard.}\}.
$$
\item
Irreducible representations of $\Cal H_{n,r}$ are parametrized by
$r$-tuples $\underline \lambda$ of size $n$.
\endroster
\endproclaim
We denote by $V_{\underline \lambda}$ the irreducible $\Cal H_{n,r}$-
module corresponding to $\underline \lambda$.
\proclaim{Proposition 1.2 ([1])}
Let $W$ be an $\Cal H_{n,r}$-module. If $\varphi_{\underline S}$
occurs in $W$ considered as a $\Cal T_{n,r}$-module,
then $W$ contains $V_{\underline \lambda}$ as an irreducible component,
where $\underline \lambda$ is the shape of $\underline S$.
\endproclaim

\heading
\S 2 The action of $\Cal H_{n,r}$ on $V^{\otimes n}$
\endheading

We will denote the basis element $v_{i_1}\otimes \cdots \otimes v_{i_n}$
of $V^{\otimes n}$ by $(i_1,\cdots , i_n)$.  Define the endomorphisms
$\theta$ and $\varpi$ on $V^{\otimes n}$ by
$$
\align
\theta (i_1,\cdots , i_n) &= (i_2,\cdots , i_n, i_1), \\
\varpi (i_1,\cdots , i_n) &= u_{i_1}q^{\mu^{(i_1)}-1}(i_1,\cdots , i_n), \\
\endalign
$$
respectively, where
$\mu^{(i)} = \#\{ j ;  1\leq j \leq n, i_j = i\}$ , the multiplicity of $i$
in the set $\{i_1,\dots ,i_n\}$.
\proclaim{Proposition 2.1}
The action of $g_i$ on $V^{\otimes n }$ defined by
$$
\align
\tilde \sigma (g_1) &= T_2^{-1} \cdots T_n^{-1}\theta \varpi, \\
\tilde \sigma (g_i) &= T_i \qquad (2 \leq i \leq n), \\
\endalign
$$
gives a representation of $\Cal H_{n,r}$.
\endproclaim

The rest of this section is devoted to proving Proposition 2.1.
We denote $T_1 = \tilde \sigma (g_1)$.
Jimbo's result [10] shows that the endomorphisms $T_i$
$(2 \leq i \leq n)$ satisfy the relations :
$$
\align
(T_i -q)(T_i +q^{-1}) &= 0 \qquad (2 \leq i \leq n), \\
T_iT_{i+ 1}T_i &= T_{i+ 1}T_iT_{i+ 1} \qquad (2 \leq i < n), \\
T_iT_j &= T_jT_i \qquad (i > j+1). \\
\endalign
$$
Hence, we only have to show
$$
\align
(T_1 -u_1)\cdots (T_1 - u_r) &= 0, \\
T_1T_2T_1T_2 &= T_2T_1T_2T_1, \\
T_1T_j &= T_jT_1 \qquad (j \geq 3). \\
\endalign
$$
We prove these relations in the following three lemmas, each of which
corresponds to each of the above relations respectively.
\proclaim{Lemma 2.2}
For $1\leq j\leq r$ and $1\leq k \leq n$, let $W_{j,k}$ be the subspace
of $V^{\otimes n}$ spanned by $\{ (i_1,\dots ,i_n) \in V^{\otimes n}
; i_k \geq j \}$, and put $W_{r+1,k} = (0)$.
\roster
\item
For $(i_1,\cdots ,i_n) \in W_{j,k}$,
we have $(i_1,\cdots ,i_k, i_{k-1},\cdot\cdot,i_n)$
 $\in W_{j,k-1}$, and
$$
T_k^{-1}(i_1,\cdots ,i_n) \equiv q^{-\delta (k)}
(i_1,\cdots ,i_k, i_{k-1},\cdots ,i_n) \qquad
\text{ mod } W_{j+1,k-1},
$$
where $\delta (k)$ is 1 or 0 according to $i_{k-1} = i_k$ or not.
In particular, we have $T_k^{-1}W_{j,k} \subset W_{j,k-1}$.
\item
For $(i_1,\cdots ,i_n) \in W_{j,1}$,
we have $(i_2,\cdots ,i_{k-1}, i_1,i_{k},\cdots ,i_n)\in W_{j,k}$, and
$$
\align
T_k^{-1}\cdots T_n^{-1}
(i_2,\cdots ,i_n,i_1)
\equiv &q^{-\mu_{i_1} (k)}
(i_2,\cdots ,i_{k-1},i_1, i_{k},\cdots ,i_n)\\
&\text{ mod } W_{j+1,k-1},
\endalign
$$
where $\mu^{(i)}(k) = \#\{ j ; k \leq j\leq n, i_j = i\}$.
\item
$(T_1 - u_j)W_{j,1} \subset W_{j+1, 1} $ for $1\leq j \leq r$.
\item
$(T_1 -u_1)\cdots (T_1 -u_r) = 0$.
\endroster
\endproclaim
\demo{Proof}
(1)
It is a direct consequence of the definition of $T_k^{-1}$.

\noindent
(2)
Use the descending induction on $k$.  The case $k = n$ is nothing but
the case $k = n$ in (1).  If
$$
\align
T_k^{-1}\cdots T_n^{-1}
(i_2,\cdots ,i_n,i_1) \equiv &q^{-\mu_{i_1} (k)}
(i_2,\cdots ,i_{k-1},i_1, i_{k},\cdots ,i_n) \\
&\text{ mod } W_{j+1,k-1},
\endalign
$$
then we have
$$
\align
T_{k-1}^{-1}T_k^{-1}\cdots T_n^{-1}
(i_2,\cdots ,i_n,i_1) \equiv & q^{-\mu_{i_1} (k-1)}
(i_2,\cdots ,i_{k-2},i_1, i_{k-1},\cdot\cdot,i_n) \\
&\text{ mod } (W_{j+1,k-2}+T_{k-1}^{-1}W_{j+1,k-1}).\\
\endalign
$$
Since $T_{k-1}^{-1}W_{j+1,k-1} \subset W_{j+1,k-2}$ by (1), the induction
proceeds.

\noindent
(3) By putting $k = 2$ in (2), we have
$$
T_2^{-1}\cdots T_n^{-1}\theta\varpi
(i_1,\cdots ,i_n) \equiv u_{i_1}
(i_1,\cdots ,i_n) \qquad
\text{ mod } W_{j+1,1}.
$$
If $i_1 >j$, then $(i_1,\cdots ,i_n)\equiv 0$ mod $W_{j+1,1}$.
Thus we have $T_1(i_1,\cdots ,i_n)$ $\in W_{j,1}$ and
$$
T_1(i_1,\cdots ,i_n) \equiv u_j(i_1,\cdots ,i_n) \qquad \text{ mod }
W_{j+1,1}.
$$
Therefore $(T_1-u_j)W_{j,1} \subset W_{j+1,1}$.

\noindent
(4) We have $(T_1-u_r)\cdots (T_1 - u_1)W_{1,1} \subset
(T_1-u_r)\cdots (T_1 - u_k)W_{k,1}$ for $1 \leq k \leq r$, which means
$(T_1-u_1)\cdots (T_1 - u_r)= 0$. \qed
\enddemo
\proclaim{Lemma 2.3}
\roster
\item
$\theta T_j = T_{j-1} \theta$ and $\varpi T_j = T_j \varpi$
for $j \geq 3$.
\item
$\theta^2 T_2 = T_n \theta^2$ and
$\theta^{-1}\varpi\theta\varpi T_2 =
T_2\theta^{-1}\varpi\theta\varpi$.
\item
$T_1T_2T_1 = (T_2^{-1}\cdots T_n^{-1})(T_2^{-1}\cdots T_{n-1}^{-1})
(\theta\varpi )^2.$
\item
We have
$$
\align
&(T_2^{-1}\cdots T_n^{-1})(T_2^{-1}\cdots T_{n-1}^{-1})T_n \\
&=(T_2^{-1}\cdots T_k^{-1})(T_2^{-1}\cdots T_{k-1}^{-1})T_k
(T_{k+1}^{-1}\cdots T_n^{-1})(T_{k}^{-1}\cdots T_{n-1}^{-1}) \\
\endalign
$$
for $k \geq 3$.
\item
$T_1T_2T_1T_2= T_2T_1T_2T_1.$
\endroster
\endproclaim
\demo{Proof}
(1) and (2) are direct consequences of the definition of $\theta$ and
$\varpi$.

\noindent
(3) By (1), we have
$$
\align
T_1T_2T_1 & = (T_2^{-1}\cdots T_n^{-1}) \theta\varpi
(T_3^{-1}\cdots T_n^{-1}) \theta\varpi \\
&= (T_2^{-1}\cdots T_n^{-1}) \theta
(T_3^{-1}\cdots T_n^{-1}) \varpi\theta\varpi \\
&=(T_2^{-1}\cdots T_n^{-1})
(T_2^{-1}\cdots T_{n-1}^{-1}) (\theta\varpi )^2. \\
\endalign
$$

\noindent
(4)  Use the descending induction on $k$.  It is obvious when $k = n$.
For $k \geq 3$, we have
$$
\gather
 (T_2^{-1}\cdots T_k^{-1})(T_2^{-1}\cdots T_{k-1}^{-1})T_k
(T_{k+1}^{-1}\cdots T_n^{-1})(T_k^{-1}\cdots T_{n-1}^{-1}) \\
= (T_2^{-1}\cdots T_{k-1}^{-1})(T_2^{-1}\cdots T_{k-2}^{-1})\times\\
(T_k^{-1}T_{k-1}^{-1}T_k)
(T_{k+1}^{-1}\cdots T_n^{-1})(T_k^{-1}\cdots T_{n-1}^{-1}) \\
= (T_2^{-1}\cdots T_{k-1}^{-1})(T_2^{-1}\cdots T_{k-2}^{-1})\times\\
(T_{k-1}T_{k}^{-1}T_{k-1}^{-1})
(T_{k+1}^{-1}\cdots T_n^{-1})(T_k^{-1}\cdots T_{n-1}^{-1}) \\
= (T_2^{-1}\cdots T_{k-1}^{-1})(T_2^{-1}\cdots T_{k-2}^{-1})
T_{k-1}
(T_{k}^{-1}\cdots T_n^{-1})(T_{k-1}^{-1}T_k^{-1}\cdots T_{n-1}^{-1}) \\
\endgather
$$
and the induction proceeds.

\noindent
(5) By putting $k = 2$ in (4), we have
$$
(T_2^{-1}\cdots T_{n}^{-1})(T_2^{-1}\cdots T_{n-1}^{-1})
T_n =
(T_{3}^{-1}\cdots T_n^{-1})(T_2^{-1}\cdots T_{n-1}^{-1}).
$$
Therefore, by (2) and (3), we have
$$
\align
T_1T_2T_1T_2 = & (T_2^{-1}\cdots T_{n}^{-1})(T_2^{-1}\cdots T_{n-1}^{-1})
(\theta\varpi )^2 T_2 \\
= & (T_2^{-1}\cdots T_{n}^{-1})(T_2^{-1}\cdots T_{n-1}^{-1})
T_n (\theta\varpi )^2  \\
= & (T_3^{-1}\cdots T_{n}^{-1})(T_2^{-1}\cdots T_{n-1}^{-1})
(\theta\varpi )^2  \\
= & T_2T_1T_2T_1. \qed
\endalign
$$
\enddemo
\proclaim{Lemma 2.4}
$T_1T_j = T_jT_1$ for $j \geq 3$.
\endproclaim
\demo{Proof}
Lemma 2.3 (1) shows that
$$
\align
T_1T_j =& T_2^{-1}\cdots T_n^{-1}\theta\varpi T_j \\
=&  T_2^{-1}\cdots T_n^{-1}T_{j-1}\theta\varpi  \\
= & T_2^{-1}\cdots T_{j-1}^{-1}T_{j}^{-1}T_{j-1}T_{j+1}^{-1}
\cdots T_n^{-1} \theta\varpi \\
= & T_2^{-1}\cdots T_{j-2}^{-1}T_{j}T_{j-1}^{-1}T_{j}^{-1}
\cdots T_n^{-1} \theta\varpi \\
= & T_jT_2^{-1}\cdots T_n^{-1}\theta\varpi \\
= & T_jT_1. \qed
\endalign
$$
\enddemo

These complete the proof of Proposition 2.1.

\TagsOnLeft
\heading
\S3 Schur-Weyl reciprocity for $(U_q(h),\Cal H_{n,r})$ on $V^{\otimes n}$
\endheading

We first give the irreducible decomposition of the representation
$\tilde\sigma$ of $\Cal H_{n,r}$ on $V^{\otimes n}$. Let
$\varLambda_{\underline 1}$ be the set of r-tuples of Young diagrams
$\underline\lambda=(\lambda^{(1)},\cdots,\lambda^{(r)})$ of size
$\sum_{i=1}^r \mid\lambda^{(i)}\mid= n$ such that each component
$\lambda^{(i)}$ has length $l(\lambda^{(i)})\le 1$. We can think
$\lambda^{(i)}$ to be a non-negative integer.

\proclaim{Theorem 3.1}
The irreducible decomposition of $V^{\otimes n}$ under the action
$\tilde\sigma$ of $\Cal H_{n,r}$ is given by
$$
V^{\otimes n}=\bigoplus_{\underline\lambda\in\varLambda_{\underline 1}}
V_{\underline\lambda}.
$$
\endproclaim
\demo{Proof} For each
$\underline\lambda\in\varLambda_{\underline 1}$, we have
$dim V_{\underline\lambda}=\frac{n!}{\prod_{i=1}^r\lambda^{(i)}!}$, and
hence the dimension of the right hand side is equal to
$$
\sum_{\underline\lambda\in\varLambda_{\underline 1}}
\frac{n!}{\prod_{i=1}^r\lambda^{(i)}!} = r^n,
$$
which coincides with $dim V^{\otimes n}$. Therefore, it suffices to
prove that $V_{\underline\lambda}\subset V^{\otimes n}$ for each
$\underline\lambda\in\varLambda_{\underline 1}$. Because of
Proposition 1.2, we only have to show that for each
$\underline\lambda\in\varLambda_{\underline 1}$, there exists a
simultaneous eigenvector for $\Cal T_{n,r}$ in $V^{\otimes n}$ with
eigenvalues $\varphi_{\underline S}$, where $\underline S$ is a certain
standard tableau of shape $\underline \lambda$.

Put $p_k=\lambda^{(r)}+\cdots+\lambda^{(k)}$ for $1\le k\le r$, and
$p_{r+1}=0$. Define $v_{\underline S}=(i_1,\cdots,i_n)$ by $i_j=k$ if
$p_{k+1}+1\le j\le p_k$.

We show that
$$
\tilde\sigma(t_j)v_{\underline S}= u_kq^{2(j-p_{k+1}-1)}v_{\underline S}
\tag 3.1
$$
if $p_{k+1}+1\le j\le p_k$. By the descending induction on $l$, we see that
$$
T_{l+1}\cdots T_{p_{k+1}+1}v_{\underline S}=
(i_1,\cdots,i_{l-1},k,i_l,\cdots,\widehat{i_{p_{k+1}+1}},\cdots,i_n)
\tag 3.2
$$
for $1\le l\le p_{k+1}$. Since $i_{p_{k+1}}=k+1>k=i_{p_{k+1}+1}$, $(3.2)$
holds for $l=p_{k+1}$. Assume $(3.2)$ for $l$. Then it is seen that
$$
T_l\cdots T_{p_{k+1}+1}v_{\underline S}=
(i_1,\cdots,i_{l-2},k,i_{l-1},i_l,\cdots,\widehat{i_{p_{k+1}+1}},\cdots,i_n)
$$
since $i_{l-1}> k$, and the induction proceeds. Putting $l=1$, we have
$$
T_2\cdots T_{p_{k+1}+1}v_{\underline S}=
(k,i_1,i_2,\cdots,\widehat{i_{p_{k+1}+1}},\cdots,i_n).
$$
Since
$$
\tilde\sigma(t_{p_{k+1}+1})=T_{p_{k+1}+2}^{-1}\cdots T_n^{-1}\theta\varpi
T_2\cdots T_{p_{k+1}+1},
$$
we have
$$
\tilde\sigma(t_{p_{k+1}+1})v_{\underline S}=
u_kq^{\lambda^{(k)}-1}T_{p_{k+1}+2}^{-1}\cdots T_n^{-1}
(i_1,\cdots,\widehat{i_{p_{k+1}+1}},\cdots,i_n,k).
$$
By a similar argument as that for $(3.2)$, we can show that
$$
\align
T_l^{-1}\cdots T_n^{-1}&(i_1,\cdots,\widehat{i_{p_{k+1}+1}},\cdots,i_n,k)\\
&=(i_1,\cdots,\widehat{i_{p_{k+1}+1}},\cdots,i_l,k,i_{l+1},\cdots,i_n)
\endalign
$$
for $p_k\le l\le n-1$. Hence we have
$$
\align
&\tilde\sigma(t_{p_{k+1}+1})v_{\underline S}= \\
&\quad u_kq^{\lambda^{(k)}-1}T_{p_{k+1}+2}^{-1}\cdots T_{p_k}^{-1}
(i_1,\cdot\cdot,\widehat{i_{p_{k+1}+1}},
\cdot\cdot,i_{p_k},k,i_{p_k+1},\cdot\cdot,i_n).
\endalign
$$
Since $i_{p_{k+1}+1}=\cdots=i_p=k$, the right hand side is equal to
$$
u_kq^{\lambda^{(k)}-1}\cdot q^{-(p_k-p_{k+1}-1)}v_{\underline S}
= u_kv_{\underline S}.
$$
This shows that $(3.1)$ holds for the case $j=p_{k+1}+1$. For the case
$p_{k+1}+2\le j\le p_k$, we only have to observe that
$T_jv_{\underline S}= qv_{\underline S}$ to see $(3.1)$. \qed
\enddemo

\bigskip
\proclaim{Theorem 3.2}
\roster
\item
Each of $\tilde\rho(U_q(h))$ and $\tilde\sigma(\Cal H_{n,r})$ is the full
centralizer algebra of the other in $End_K(V^{\otimes n})$.
\item
For $\underline\lambda=(\lambda^{(1)},\cdots,\lambda^{(r)})\in
\varLambda_{\underline 1}$,
$$
V_{\underline\lambda}=\{v\in V^{\otimes n}\mid
\tilde\rho(q^{\epsilon_i})v=q^{\lambda^{(i)}}v\;\text{for\;}1\le i\le r \}
$$
\endroster
\endproclaim
\demo{Proof} We first observe that $\tilde\rho(q^{\epsilon_i})$ commutes
with $T_1,\cdots,T_n$ for $1\le$ $i\le r$, which easily follows from
$$
\tilde\rho(q^{\epsilon_i})(i_1,\cdots,i_n)=q^{\mu^{(i)}}(i_1,\cdots,i_n),
$$
and the fact that $T_1,\cdots,T_n$ preserve the weight
$\underline\mu\in\varLambda_{\underline 1}$ of the vector $(i_1,\cdots,i_n)$.

Let $\phi_{\underline\mu}$ be the character of $U_q(h)$ defined by
$$
\phi_{\underline\mu}(q^{\epsilon_i})=q^{\mu^{(i)}}\quad(1\le i\le r)
$$
Then we have the isotypical decomposition of $V^{\otimes n}$ as
$U_q(h)$-modules;
$$
V^{\otimes n}=\sum_{\underline\mu}(V^{\otimes n})_{\underline\mu},
\tag 3.3
$$
where
$$
(V^{\otimes n})_{\underline\mu}
=\{v\in V^{\otimes n}\mid \tilde\rho(q^{\epsilon_i})v=
q^{\mu^{(i)}}v \quad\text{for\;}1\le i\le r \}.
$$
It is apparant that the characters $\phi_{\underline\mu}$ and
$\phi_{\underline\mu'}$ are inequivalent for distinct weights
$\underline\mu$ and $\underline\mu'$. Therefore, dimension
of $\tilde\rho(U_q(h))$ is equal to $\mid\varLambda_{\underline 1}\mid$.

On the other hand, by Schur's lemma, the dimension of the commu\-tant
$\tilde\sigma(\Cal H_{n,r})'$ of $\tilde\sigma(\Cal H_{n,r})$
is equal to $\mid\varLambda_{\underline 1}\mid$.
Since $\tilde\rho(U_q(h))$ $\subset$ $\tilde\rho(\Cal H_{n,r})'$, we can
conclude that they coincide.

Let us recall that for any simple algebra acting on a module say, $M$,
its image in $End_K(M)$ is isomorphic to the direct product of matrix
algebras which correspond to irreducible representations occurring in $M$.
Thereby we know that
$$
\tilde\sigma(\Cal H_{n,r})\simeq\bigoplus_{\underline\lambda
\in\varLambda_{\underline 1}}End_K(V_{\underline\lambda}).
$$
Hence $dim \tilde\sigma(\Cal H_{n,r})=\sum_{\underline\lambda\in
\varLambda_{\underline 1}}
\binom n{\lambda^{(1)},\cdots,\lambda^{(r)}}^2$.

On the other hand, the decomposition $(3.3)$ shows that the dimension of
the commutant $\tilde\rho(U_q(h))'$ of $\tilde\rho(U_q(h))$ is equal to
$$
\sum_{\underline\mu\in\varLambda_{\underline 1}}
\binom n{\mu^{(1)},\cdots,\mu^{(r)}}^2.
$$
Therefore,
we have $\tilde\rho(U_q(h))'=\tilde\sigma(\Cal H_{n,r})$.

Since the space $(V^{\otimes n})_{\underline\lambda}$ has the element
$(r,\cdot\cdot,r,r-1,\cdots\cdot,1)$ in which
each $k$ repeats $\lambda^{(k)}$ times, the space contains
$V_{\underline\lambda}$. Combining Theorem 3.1 and $(3.3)$, we see
$V_{\underline\lambda}=(V^{\otimes n})_{\underline\lambda}$ for any
$\underline\lambda\in\varLambda_{\underline 1}$ as desired. \qed
\enddemo

\heading
\S4 The action of certain lattices of $U_q(h)$ on $V^{\otimes n}$
\endheading

If we put aside the representation theory, the story we have considered
in previous sections can be discussed not only over fields but also over
$A=\Bbb Z[q,q^{-1},u_1,\cdots,u_r]$.

Let us denote by $V_A^{\otimes n}$ the $A$-lattice spanned by the basis
elements $(i_1,\cdots,i_n)$. The Hecke algebra
$\Cal H_{n,r}$ has a natural $A$-lattice which is the $A$-subalgebra
generated by $g_1,\cdots,g_n$. It is obvious that
$\tilde\sigma(\Cal H_{n,r}(A))$ preserves $V_A^{\otimes n}$.

There are several natural ways to choose an $A$-lattice in $U_q(h)$.
We consider two lattices, one is the $A$-subalgebra generated by
$q^{\epsilon_i}$'s, and the other is analogous to the Cartan part of
the Kostant $\Bbb Z$-form introduced by G.Lusztig [12].
These are mapped to $A$-subalgebras in $End_A(V_A)$ by $\tilde\rho$,
which are $A$-free of finite rank.
In the following, we give $A$-free bases of these, as well as proving
that these preserve $V_A^{\otimes n}$.

Let $\Cal S$ be a set of dominant weights in an alcove as follows.
$$
\Cal S=\{\underline\nu=(\nu_1,\cdots,\nu_{r-1})\mid n\ge\nu_{1}\ge
\cdots\ge\nu_{r-1}\ge0\}.
$$
Put, $\nu_{r-1}=p_r,\;\nu_{i-1}-\nu_i=p_i\quad(2\le i< r),\;n-\nu_1=p_r$.
Note that this bijection $\underline\nu\leftrightarrow\underline
p=(p_1,\cdots,p_r)$ shows that the cardinality of $\Cal S$ is equal to
the dimension of $\tilde\rho(U_q(h))$.

We define a polynomial of $(r+1)$-variables as follows.
$$
\align
F_{\underline\nu}(X_0,\cdots,X_r)=
&(X_0-X_1)(X_0-qX_1)\cdots\cdots(X_0-q^{p_1-1}X_1)\\
&(X_1-X_2)\cdots\cdots(X_1-q^{p_2-1}X_2)\\
&\qquad\qquad\cdots\cdots\cdots\cdots\\
&(X_{r-1}-X_r)\cdots(X_{r-1}-q^{p_r-1}X_r),
\endalign
$$
which is homogeneous of degree $n$. In the following,
we are mainly concerned with $F_{\underline\nu}(q^nX_r,X_1,\cdots,X_r)$.

\proclaim{Proposition 4.1}
\roster
\item
$$
\text{If} \qquad F_{\underline\nu}(q^nX_r,X_1,\cdots,X_r)=
\sum_{\underline\mu\in\varLambda_{\underline 1}}
a_{\underline\mu\;\underline\nu}X_1^{\mu^{(1)}}\cdots X_r^{\mu^{(r)}},
$$
then, for any $\underline\nu'$, $\underline\nu$ in $\Cal S$,
$$
\sum_{\underline\mu\in\varLambda_{\underline 1}}
q^{(\underline\nu',\underline\mu)}
a_{\underline\mu\;\underline\nu}=
\delta_{\underline\nu',\underline\nu}F_{\underline\nu}
(q^n,q^{\nu_1},\cdots,q^{\nu_{r-1}},1),
$$
where, $q^{(\underline\nu',\underline\mu)}$ stands for
$q^{\nu'_1\mu^{(1)}+\cdots+\nu'_{r-1}\mu^{(r-1)}}$.

\item
For any $\underline \alpha=(\alpha_1,\cdots,\alpha_r)\in\Bbb Z^r$, we have
$$
\align
\tilde\rho(&q^{\alpha_1\epsilon_1+\cdots+\alpha_r\epsilon_r})=\\
&\sum_{\underline\nu\in T}
\frac{F_{\underline\nu}(q^{n+\alpha_r},q^{\alpha_1},\cdots,
q^{\alpha_{r-1}},q^{\alpha_r})}
{F_{\underline\nu}(q^n,q^{\nu_1},\cdots,q^{\nu_{r-1}},1)}
\tilde\rho(q^{\nu_1\epsilon_1+\cdots+\nu_{r-1}\epsilon_{r-1}}).
\endalign
$$
\item
$$
\frac{F_{\underline\nu}(q^{n+\alpha_r},q^{\alpha_1},\cdots,
q^{\alpha_{r-1}},q^{\alpha_r})}
{F_{\underline\nu}(q^n,q^{\nu_1},\cdots,q^{\nu_{r-1}},1)}
\in \Bbb Z[q,q^{-1}].
$$
\item
The $A$-algebra generated by $\tilde\rho(q^{\epsilon_i})$ $(1\le i\le r)$
has an $A$-free basis  \linebreak
$\{\tilde\rho(q^{\nu_1\epsilon_1+\cdots+\nu_{r-1}\epsilon_{r-1}})
\mid \underline\nu\in \Cal S\}$.
\endroster
\endproclaim
\demo{Proof} (1) We will evaluate the left hand side, which is equal to

\noindent
$F_{\underline\nu}(q^n,q^{\nu'_1},\cdots,q^{\nu'_{r-1}},1)$.
Assume that it is nonzero. Then we have
$$
\align
\nu'_{r-1}&\ne 0,1,\cdots,p_r-1,\\
\nu'_{r-2}&\ne \nu'_{r-1},\nu'_{r-1}+1,\cdots,\nu'_{r-1}+p_{r-1}-1,\\
&\cdots\cdots\\
\nu'_1&\ne \nu'_2,\nu'_2+1,\cdots,\nu'_2+p_2-1,\\
n&\ne \nu'_1,\nu'_1+1,\cdots,\nu'_1+p_1-1,
\endalign
$$
which leads to
$$
\gather
\nu'_{r-1}\ge p_r,\;\nu'_{r-2}-\nu'_{r-1}\ge p_{r-1},\cdots,
\nu'_1-\nu'_2\ge p_2,\\
n-\nu'_1\ge p_1.
\endgather
$$
Since the sum of these on both sides is $n$, these inequalities must be
equalities. It deduces $\underline\nu'=\underline\nu$, which proves (1).

\noindent
(2) A conclusion of (1) is that the matrix
$\bigl(q^{\nu_1\mu^{(1)}+\cdots+\nu_{r-1}\mu^{(r-1)}}\bigr)$ whose
rows are indexed by $\nu$ and columns by $\mu$, is non-singular,
which means that  \linebreak
$\{\tilde\rho(q^{\nu_1\epsilon_1+\cdots+\nu_{r-1}\epsilon_{r-1}})
\mid \underline\nu\in \Cal S\}$
is a $K$-basis of $\tilde\rho(U_q(h))$. Hence we can write
$$
\tilde\rho(q^{\alpha_1\epsilon_1+\cdots+\alpha_r\epsilon_r})=
\sum_{\underline\nu\in \Cal S}
b_{\underline \alpha\underline\nu}
\tilde\rho(q^{\nu_1\epsilon_1+\cdots+\nu_{r-1}\epsilon_{r-1}})
$$
for any fixed $\underline \alpha=(\alpha_1,\cdots,\alpha_r)\in\Bbb Z^r$.
If we apply both sides to a weight vector of $V^{\otimes n}$ with weight
$\underline\mu$, we have
$$
q^{\alpha_1\mu^{(1)}+\cdots+\alpha_r\mu^{(r)}}=
\sum_{\underline\nu\in \Cal S}
b_{\underline \alpha\underline\nu}
q^{\nu_1\mu^{(1)}+\cdots+\nu_{r-1}\mu^{(r-1)}}.
$$
Hence,
$$
\align
F_{\underline\nu}(q^{n+\alpha_r},q^{\alpha_1},\cdots,q^{\alpha_r})&=
\sum_{\underline\mu}
a_{\underline\mu\;\underline\nu}
q^{\alpha_1\mu^{(1)}+\cdots+\alpha_r\mu^{(r)}}\\
&=\sum_{\underline\mu,\underline\nu'}
a_{\underline\mu\;\underline\nu}b_{\underline \alpha\underline\nu'}
q^{\nu'_1\mu^{(1)}+\cdots+\nu'_{r-1}\mu^{(r-1)}}\\
&=\sum_{\underline\nu'}
b_{\underline \alpha\underline\nu'}
\delta_{\underline\nu'\underline\nu}
F_{\underline\nu}(q^n,q^{\nu_1},\cdots,q^{\nu_{r-1}},1)\\
&=b_{\underline \alpha\underline\nu}
F_{\underline\nu}(q^n,q^{\nu_1},\cdots,q^{\nu_{r-1}},1),
\endalign
$$
by which we have the required coefficients.

\noindent
(3) It is obvious since the quotient
$$
\frac{F_{\underline\nu}(q^{n+\alpha_r},q^{\alpha_1},\cdots,
q^{\alpha_{r-1}},q^{\alpha_r})}
{F_{\underline\nu}(q^n,q^{\nu_1},\cdots,q^{\nu_{r-1}},1)}
$$
is a product of $q$-binomial coefficients up to a power of $q$.

\noindent
(4) This is a direct consequence of (2) and (3). \qed
\enddemo

The next $A$-lattice we consider is the $A$-subalgebra of $U_q(h)$
generated by $q^{\epsilon_i}$'s and
$$
\bmatrix q^{\pi} \\ N \endbmatrix
=\prod_{s=1}^N\frac{q^{\pi-s+1}-q^{-\pi+s-1}}{q^s-q^{-s}},
$$
where $N\ge0$, and $\pi\in P=\Bbb Z\epsilon_1+\cdots+\Bbb Z\epsilon_n$.

If $\pi$ is replaced by an integer $l$, it is a $q$-binomial coefficient,
which we denote by $\bmatrix l \\ N\endbmatrix$.

For comparison, we recall Lusztig's Kostant $\Bbb Z$-form for $U_q(sl_r)$
([12]). It is the $\Bbb Z[q,q^{-1}]$-algebra generated by
$e_i^{(N)}=e_i^N/[N]!$, $f_i^{(N)}=f_i^N/[N]!$, where
$[N]!=\prod_{s=1}^N\frac{q^s-q^{-s}}{q-q^{-1}}$. The Cartan part of
this algebra has a $\Bbb Z[q,q^{-1}]$-free basis
$$
\bigl\{ \prod_{i=1}^{r-1}\bigl(K_i^{\delta_i}
\bmatrix K_i \\ t_i\endbmatrix \bigr)
\mid t_i\ge0,\;\delta_i=0,1 \bigr\},
$$
where $K_i=q^{\epsilon_i-\epsilon_{i+1}}$ $(1\le i< n)$ and
$$
\bmatrix K_i \\ N \endbmatrix
=\prod_{s=1}^N\frac{q^{-s+1}K_i-q^{s-1}K_i^{-1}}{q^s-q^{-s}}.
$$

\proclaim{Proposition 4.2}
\roster
\item
Put
$
\bmatrix q^{\epsilon}\\ \underline\nu \endbmatrix
=\prod_{i=1}^r\bmatrix q^{\epsilon_i}\\ \nu^{(i)} \endbmatrix
$
for $\underline\nu\in\varLambda_{\underline 1}$. Then we have,
$$
\tilde\rho\bigl(\bmatrix q^{\epsilon}\\ \underline\nu\endbmatrix\bigr)
(i_1,\cdots,i_n)
=\delta_{\underline\mu,\underline\nu}(i_1,\cdots,i_n),
$$
where $\underline\mu$ is the weight of $(i_1,\cdots,i_n)$.
\item
For $\underline\alpha=(\alpha_1,\cdots,\alpha_r)\in\Bbb Z^r$,
$\underline N=(N_1,\cdots,N_r)\in\Bbb Z^r$ and $\pi_i\in P$, we have
$$
\tilde\rho\bigl(\prod_{i=1}^r(q^{\alpha_i\epsilon_i}
\bmatrix q^{\pi_i} \\ N_i\endbmatrix)\bigr)
=\sum_{\underline\nu\in\varLambda_{\underline 1}}
\bigl(\prod_{i=1}^r q^{\alpha_i\nu^{(i)}}
{\bmatrix (\pi_i,\underline\nu)\\ N_i\endbmatrix}
\bigr)
\tilde\rho\bigl(\bmatrix q^{\epsilon}\\ \underline\nu\endbmatrix \bigr).
$$
\item
The $A$-algebra generated by $\tilde\rho(\bmatrix q^{\pi}\\ N \endbmatrix)$
and
$\tilde\rho(q^{\pi})$ $(\pi\in P,\;N\ge0)$ has an $A$-free basis
$
\{\tilde\rho\bigl(\bmatrix q^{\epsilon}\\ \underline\nu\endbmatrix\bigr)\mid
\underline\nu\in\varLambda_{\underline 1}\}.
$
\endroster
\endproclaim
\demo{Proof}(1) It is enough to evaluate $\bmatrix \nu_i\\ \mu_i\endbmatrix$.
If all of these are nonzero, then we must have $\mu_i\ne 0,1,\cdots,\nu_i-1$
for all $i$. But it is nothing but $\mu_i\ge\nu_i$, and we conclude that
$\underline\mu=\underline\nu$, in which case the value is $1$.

(2) is a direct consequence of (1). Then (3) easily follows since
$q$-binomial coefficients are Laurent polynomials in $q$. \qed
\enddemo


\heading
Appendix
\endheading
We explain a finite field version of the classical Schur-Weyl reciprocity
in our setting.

Let $p$ be an odd prime such that $r\ge n$, $p\ge r+n$ and $r$ divides $p-1$.
Set $q=p^r$ and let $\Bbb F_q$ be the field of $q$ elements.
The general linear group over the field $\Bbb F_q$, which we denote by
$G=GL(r,q)$, admits the Frobenius actions
$$
F_1((g_{ij}))=(g_{ij}^p),\;\text{and}\;
F_w((g_{ij}))=w^{-1}(g_{ij}^p)w\quad\text{for $g=(g_{ij})$},
$$
where $w=E_{r,1}+\sum_{i=1}^{r-1}E_{i,i+1}$.
We denote by $G^{F_w}$ the group of Frobenius fixed points with respect to
$F_w$, which is isomorphic to $GL(r,p)$.

We are dealing with the following $G$-module:
$$
E=V\otimes_{\Bbb F_q}V^{F_w}\otimes_{\Bbb F_q}\cdots\otimes_{\Bbb F_q}
V^{F_w^{n-1}},
$$
where $V=\Bbb F_q^r$ is the natural representation of $G$, and
$$
\hat\rho(g)(v_{i_1}\otimes\cdots\otimes v_{i_n})=
gv_{i_1}\otimes F_w(g)v_{i_2}\otimes\cdots\otimes F_w^{n-1}(g)v_{i_n}
$$
Steinberg's tensor product theorem [16] says that $E$ is isomorphic to the
irreducible module with highest weight $(1+\cdots+p^{r-1})\epsilon_1$
as a $SL(n,q)$-module.($\epsilon_1$ is the highest weight of $V$.)

Note that the set of irreducible representations of $SL(r,\overline{\Bbb F_q})$
 with highest weights lying in an alcove of the weight lattice exhausts
all irreducible representations of $SL(r,q)$.

As a $G^{F_w}$-module, $E$ is nothing but $V^{\otimes n}$, which admits
$\frak S_n$-action as before. We also denote by $\rho$ and $\sigma$
the action of $G^{F_w}$ and $\frak S_n$ on $V^{\otimes n}$ respectively.

\bigskip
\proclaim{Lemma}
Each of $\sigma(\Bbb F_q\frak S_n)$ and $\rho(\Bbb F_q G^{F_w})$ is the
full centralizer algebra of the other.
\endproclaim
\demo{Proof}
Let us review results in [5]. Let $\lambda=(\lambda_1,\cdots,\lambda_n)$
be a Young diagram of size $n$. Then, by definition, the Weyl module
$W_{\lambda}$ is a cyclic module generated by a highest weight vector
$\varPhi$ in $V^{\otimes n}$. (For its definition, see [5;4.2])
This module can be defined over $\Bbb F_q$. Besides, the collection of
$\frak S_n$ translations of $\varPhi$ spans the Specht module $S_{\lambda}$.
Since all Young diagrams are $p$-regular under our assumption, these
Specht modules are irreducible.

Corollary 2 of [5,p.232] states that $W_{\lambda}$ is absolutely
irreducible if and only if $\lambda_1\ge\lambda_2\ge\cdots\ge\lambda_n$,
$\lambda_1-\lambda_n+n\le p$. Hence, under our assumption, $W_{\lambda}$
is irreducible, and the space of highest weights in the $W_{\lambda}$-
isotypical component of $V^{\otimes n}$ apparantly contains translation
of $\varPhi$ under $\frak S_n$-action. This shows that $V^{\otimes n}$
contains the direct sum of $W_{\lambda}\otimes S_{\lambda}$'s.
By counting dimensions, we have
$$
V^{\otimes n}=\bigoplus_{\lambda}W_{\lambda}\otimes S_{\lambda}.
$$
Thereby, $V^{\otimes n}$ is a multiplicity free semi-simple
$G^{F_w}\times \frak S_n$-module, which proves that each of
$\sigma(\Bbb F_q\frak S_n)$
and $\rho(\Bbb F_q G^{F_w})$ is the full centralizer algebra of the other.
\qed
\enddemo

Let $F_1$ be the natural Frobenius action on $V$.
If we set $s_1=F_1\otimes id_V\otimes\cdots\otimes id_V$,
then one can extend the action of $\frak S_n$ on $V^{\otimes n}$ to
that of $\frak S_{n,r}$. We denote this action of $\frak S_{n,r}$ by
$\tilde\sigma$.

Put $T_0=G^{F_1}\cap G^{F_w}$. Each element of $T_0$ is of the form
$$
h\bigl(diag(\Bbb F_p^{\times},\cdots,\Bbb F_p^{\times})\bigr)h^{-1},
$$
where $h=\sum \gamma^{(i-1)(j-1)}E_{ij}$, $\gamma$ is an element in
$\Bbb F_p^{\times}$ whose order is precisely $r$. Thus $T_0$ is a split
torus of $G^{F_w}$. By restricting $\rho$ to $T_0$, we have the action
of $T_0$ on $V^{\otimes n}$, which we denote by $\tilde\rho$.

\proclaim{Proposition}
Each of $\tilde\sigma(\Bbb F_q\frak S_{n,r})$ and $\rho(\Bbb F_qT_0)$ is
the full centralizer algebra of the other.
Hence, correponding to the inclusion $\hat\rho(G) \supset
\rho(G^{F_w}) \supset \tilde\rho(T_0)$, we have their
full centralizer algebras
$\Bbb F_q \subset \sigma(\Bbb F_q\frak S_n) \subset
\tilde\sigma(\Bbb F_q\frak S_{n,r})$ on $E$.
\endproclaim
\demo{Proof} As we have remarked in $\S4$, each irreducible component of
$V_A^{\otimes n}$ is stable under $\Cal H_{n,r}(A)$-action. On the other
hand, we know that its reduction modulo $u_i=q^{i-1}$ and
$q=e^{2\pi\sqrt{-1}/r}$ remains irreducible by a result proved in [2].
Hence we have
$$
E=\bigoplus_{\underline\lambda\in\varLambda_{\underline 1}}
V_{\underline\lambda}.
\tag A.1
$$
If we choose columns of the matrix $h$ as a basis for $V$, these are
simultaneous eigenvectors for $T_0$ correponding to distinct weights of
$T_0$. Hence $E$ is decomposed as $T_0$-module into
$\bigoplus_{\mu\in\varLambda_{\underline 1}}(E)_{\mu}$, such that each weight
space is nonzero and $\Bbb F_q\frak S_{n,r}$-stable, since one can easily
check that $F_1$ and $\tilde\rho(T_0)$ commute on $V$.

 Comparing this weight decomposition with (A.1), we have that these weight
spaces are irre\-ducible $\Bbb F_q\frak S_{n,r}$
-module. Therefore, $V^{\otimes n}$ is a multiplicity
-free semi
-simple $T_0\times \frak S_{n,r}$
-module, by which we conclude the required result.
\qed
\enddemo


\heading
List of Symbols
\endheading
$\frak S_n$, $\frak S_{n,r}$, $U_q(h)$, $U_q(gl_m)$, $U_q(gl_r)$, $U(\frak g)$
$t_i$, $T_i$


\enddocument